\documentclass[aps,pre,twocolumn,groupedaddress,superscriptaddress,showpacs]{revtex4-1}
\usepackage{amsmath}
\usepackage{graphicx}
\usepackage{float}
\usepackage{color}

\begin{document}
  \bibliographystyle{unsrt}
  \title {Analytical and numerical study of particles with binary adsorption}

  \author{C. S. Dias}
   \email{cristovao@fisica.uminho.pt}
    \affiliation{GCEP-Centro de F\'isica da Universidade do Minho, 4710-057 Braga, Portugal}

  \author{N. A. M. Ara\'ujo}
   \email{nuno@ethz.ch}
    \affiliation{Computational Physics for Engineering Materials, IfB, ETH Z\"{u}rich, Schafmattstr. 6, CH-8093 Z\"{u}rich, Switzerland}

  \author{A. Cadilhe}
   \email{cadilhe@lanl.gov}
    \affiliation{Theoretical Division, MSK 717, Los Alamos National Laboratory, Los Alamos, NM 87545, USA}

  \begin{abstract}

Electro-oxidation of ethanol represents a key process in fuel-cells technology. We introduce a generalization of the random sequential adsorption model to study the long timescale and large length scale properties of the electro-oxidation process. We provide an analytical solution for one dimension and Monte Carlo results in two dimensions. We characterize the coverage and percolation properties of the jammed state and unveil the influence of quenched impurities in the selectivity of oxidation products.

  \end{abstract}
  \maketitle

\section{Introduction}
 Ethanol has been considered for fuel cells given its low toxicity and abundance \cite{Wang2004}. The electro-oxidation of ethanol on the surface of catalysts can follow multiple reaction pathways leading to several different products, which are strongly affected by, e.g., concentration, presence of impurities, and temperature. In this work, we introduce a model, inspired on the random sequential adsorption model (RSA) of dimers, to analyze properties of the oxidation process such as their dependence on the binding configuration, binding rates, and reaction pathway probabilities.

 Ethanol electro-oxidation has recently been studied through density function theory \cite{Wang2008a,Wang2007}, providing possible reaction pathways for the adsorption and catalysis of ethanol.
 However, the time-dependence of the coverage based on the various reaction pathways would become computationally prohibitive using density functional theory.
 Nonetheless, in the limit of low mobility of bound reaction products, large systems sizes, and long timescales a study based on the adoption of a square lattice to represent the (100) substrate for the various reaction products becomes appropriate.
 In this limit, ethanol electro-oxidation can be described as adsorption of a dimer on the substrate, thus occupying two adjacent lattice sites, as it cleaves \cite{Wang2008a,Wang2007} or, in the presence of neighboring pre-adsorbed species, adsorb as a monomer. A key feature of the model is to provide a configuration dependent rule for the landing site and study the influence on the adsorption rates of the oxidation process. We also study the influence of immobile impurities and understand their role in achieving selectivity of adsorbed species. Our model can also  be extended to analyze other cleavage mechanisms like, e.g., the one of sugars \cite{Parpot2006}.

 The random sequential adsorption model (RSA) has been utilized to describe adsorption in the limit of low surface mobility and of negligible desorption rate \cite{Evans1993a, Cadilhe2007, Araujo2006, Araujo2010b,Bonnier2001,Bonnier1994,Privman1994,Brosilow1991,Feder1980a,Araujo2010}. Adsorption attempts occur sequentially at randomly selected sites, where particles solely interact through excluded volume. Generalized versions have been proposed where the rates of adsorption are dependent on the local configuration \cite{Evans1993a,Evans1985}, which might, e.g., explain the selectivity of adsorbed species \cite{Lopez2008}. Further extensions have been considered to study a wide range of problems, such as chemical reactions \cite{Gonz1974, Flory1939,McLeod1999}, adsorption on membranes \cite{Finegold1979}, as well as protein and colloid adsorption \cite{Feder1980,Onoda1986} with and without pre-adsorbed impurities \cite{RamirezPastor2000,Zuppa1999,Stacchiola1998,Kondrat2006, Lee1996, Bennaim1994}.

 In this paper, we study the model above delineated both in one and two dimensions. In the one-dimensional study, we were able to establish a closed hierarchy of rate equations for which we could obtain exact, closed form solutions. To complement and extend the insight provided by this approach, we also performed a Monte Carlo based study for the relevant two-dimensional case.

 The paper is organized as follows: in the next section we introduce the model, while in Section~\ref{sec:analytical} an analytical derivation is exactly solved in three specific limits. Monte Carlo simulations extend the one-dimensional results to the more realistic case of a substrate as described in section~\ref{sec:MC_1d} with results provided in section~\ref{sec:2d_results} for substrates with and without impurities. Final remarks are presented in section~\ref{sec:conclusion}.

\section{Model}\label{sec:model}

 Ethanol oxidation is of great relevance to the society, since each molecule releases twice as much energy as one methanol molecule \cite{Farias2007}, posing it as a candidate to replace several sources of energy \cite{Lynd1996}. Recently, Wang and Liu \cite{Wang2008a} proposed a mechanism for ethanol electro-oxidation on Pt(100) and Pt(111) substrates, which can be summarized in three pathways \cite{Wang2007}: 
  \begin{enumerate}
    \item The \textit{OH path}, where the cleavage of the hydroxyl group leads to the formation of acetaldehyde which is then adsorbed; 
    \item The \textit{CH path} where $CH_3CHOH$ is an intermediate product that degrades into $CH_3COH$; 
    \item The \textit{concerted path} where the ethanol molecule looses two hydrogens followed by the desorption of acetaldehyde.
  \end{enumerate}

 Wang and Liu have shown that for the Pt(100) surface (which can be mapped onto a square lattice) the relevant pathway is mainly the \textit{CH path} \cite{Wang2008a}, where ethanol adsorption leads to the formation of acetyl ($CH_3CO$). The work of Wang and Liu \cite{Wang2008a} discloses that the adsorption mechanism is strongly influenced by the actual surface coverage. At low surface coverages, the acetyl dehydrogenates into $CH_2CO$ or $CHCO$, which leads to a $C-C$ bond cleavage, yielding $CH$ and $CO$ fragments. At oxidative conditions, both fragments react with the oxygen, $O$, present on the surface and desorb as carbon dioxide $CO_2$. Desorption of cleaved products can be neglected for non-oxidative conditions, which leads to a jammed state, whereas, when the surface coverage is high, the $C-C$ bond cleavage is blocked, and the surface becomes poisoned by acetyl.
 
 Based on the proposed mechanism, we introduce a model which can be summarized by the following rules,
   \begin{align}
    A + 2v \stackrel{k_d}{\longrightarrow} 2B \nonumber \\
    A + v \stackrel{k_m}{\longrightarrow} C,
    \label{eq.adsorption_mechanism}
   \end{align}
where $A$ represents ethanol, $v$ an empty site, $2B$ represents cleaved products, $C$ is acetyl, and $k_d$ ($k_m$) stands for dimer (monomer) production rates. Note that, as discussed below, adsorption as a monomer (with product $C$) can only occur in the neighborhood of an occupied site.

 \begin{figure}[t]
   \begin{center}
    \includegraphics[width=8cm]{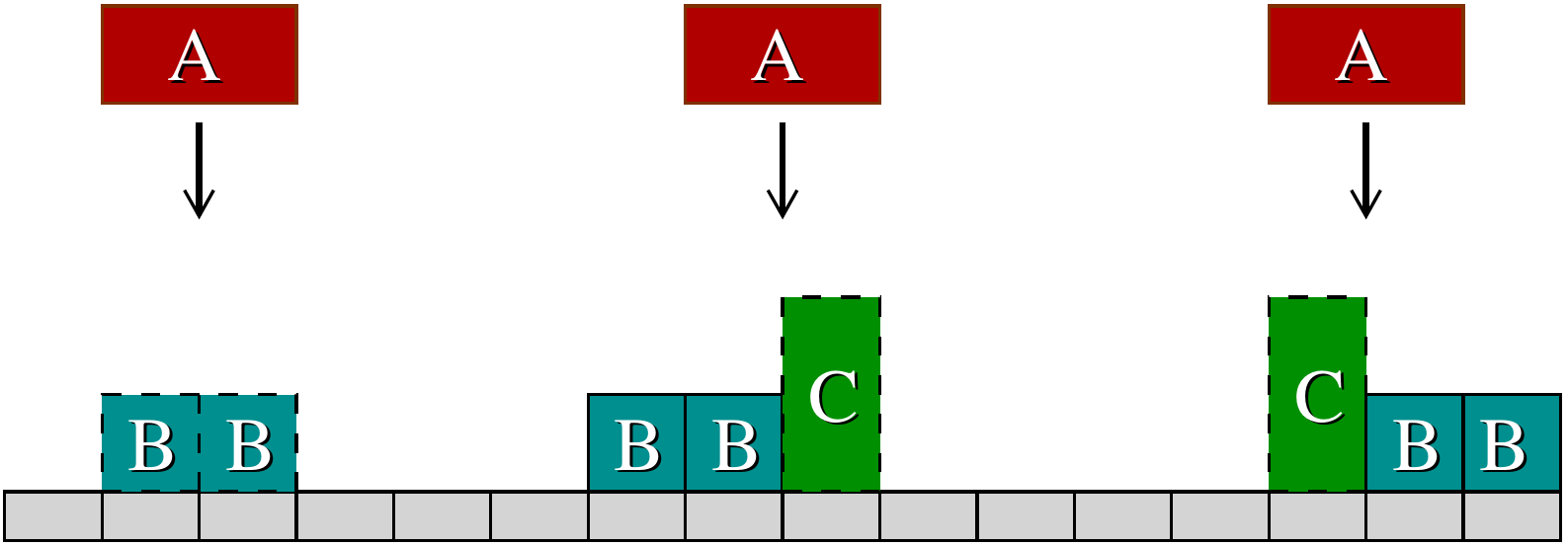}\\
    \caption{Schematic representation of the adsorption rules (1). The
A-species (red) can deposit either as dimers, on two empty sites,
yielding B-products (blue), or as monomers, on one empty site with at
least one occupied neighbor, yielding C-products (green). In the ethanol
oxidation, $A$ is the ethanol, $B$ stands for the
two products ($CH$ and $CO$), and $C$ the acetyl.}
    \label{fig.model_rules}
   \end{center}
   \end{figure}

 As cartooned in Fig.~\ref{fig.model_rules}, dimers are uniformly formed on the substrate (lattice). Successful adsorption of dimers requires two neighboring empty sites. If only one is available, the species adsorbs as a monomer. When both sites are occupied, due to the excluded volume interaction, the adsorption attempt fails and the particle attempting adsorption is no longer considered. The model differs from traditional cooperative sequential adsorption models \cite{Evans1993a} since, for the latter, the rates of adsorption depend on the state of the nearest-neighbors but not, as in the present model, on the occupation of the local configuration provided by neighboring adsorbed sites. Despite the focus on the electro-oxidation of ethanol, the model could be utilized for the study of any other process dependent on the local configuration rather than on constant, configuration independent, deposition rates.

\section{Analytical Study}\label{sec:analytical}

  The time dependence of the coverage and distribution of empty sites can be analytically obtained by establishing a closed hierarchy of rate equations as explained below. To account for different rates for monomer and dimer adsorption (see Eq.~\ref{eq.adsorption_mechanism}), we consider a competitive deposition of monomers and dimers, with different deposition rates ($k_m$ and $k_d$, respectively), under the constraint that monomers can only deposit in the neighborhood of occupied sites. Results are divided into three limiting cases: equal rates for dimers and monomers (Section \ref{sec:eq_abs}), preferential dimer site adsorption (Section \ref{sec:pref_carb}), and different deposition rates for monomers and dimers (Section \ref{sec:diff_rates}).

\begin{figure}[t]
   \begin{center}
    \includegraphics[width=8cm]{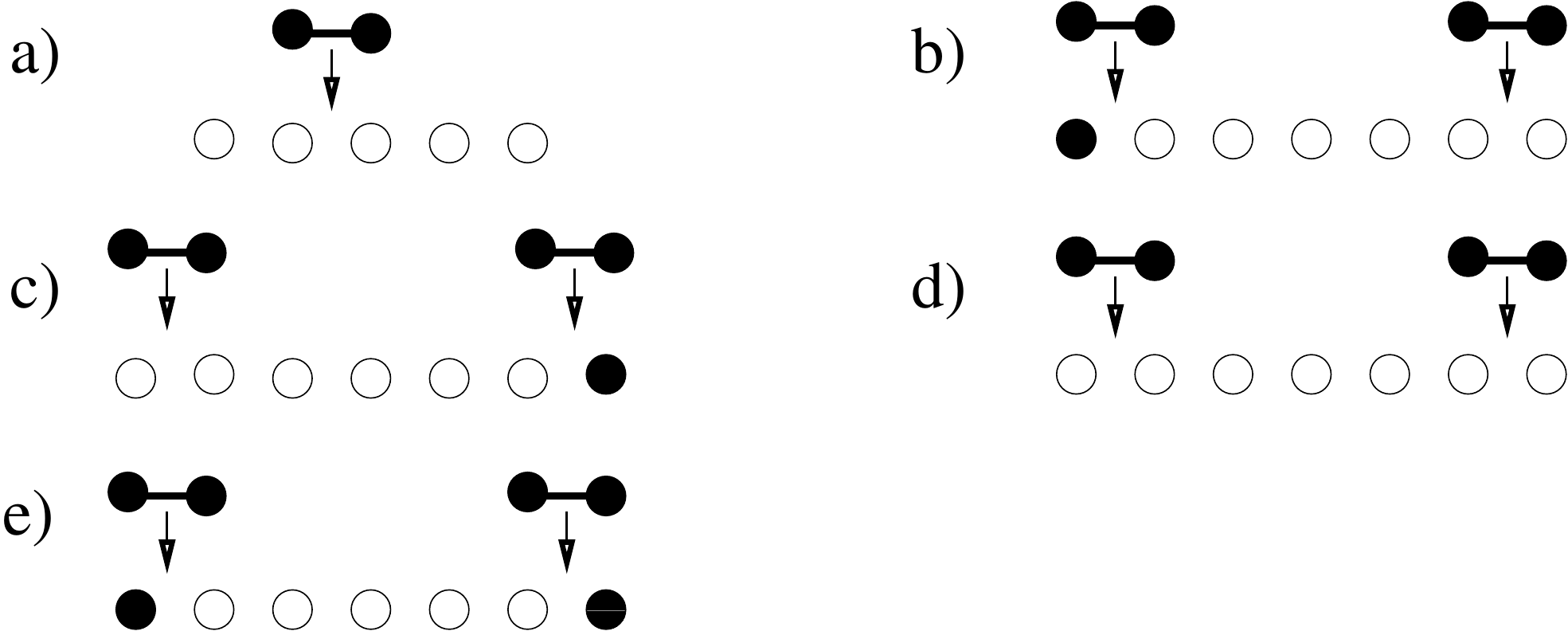}\\
    \caption{Catalog of possible adsorption attempts of a dimer on a segment of empty sites with length $n=5$, which is part of a larger segment with length $\ell$ ($\ell\geq n$). For adsorption events with both landing sites in the segment $n$, a dimer is adsorbed (a). For adsorption events with a single landing site in the $n$ segment either a dimer or a monomer is adsorbed, depending on the occupation state of the neighboring site (b, c, d, and e).
    \label{fig.anal_rules}}
   \end{center}
   \end{figure}

  Let us start by considering a segment of empty sites with size $n$ which is part of a larger (or equal) one with size $\ell\geq n$. Since the neighboring sites of this segment are not necessarily occupied, the possible adsorption events depend on the configuration of the neighbors. A segment can be reduced in size by adsorption of a monomer on the left-hand side of the segment (Fig.~\ref{fig.anal_rules}b), the right-hand side (Fig.~\ref{fig.anal_rules}c), or on both sides (Fig.~\ref{fig.anal_rules}e). It can also be split (or reduced in size) by the adsorption of a dimer on any site of the segment (Fig.~\ref{fig.anal_rules}a), the right-hand side (Fig.~\ref{fig.anal_rules}b), the left-hand side (Fig.~\ref{fig.anal_rules}c), or on both sides (Fig.~\ref{fig.anal_rules}d). 

  We define $P_n$ as the probability that a randomly selected site belongs to any segment $n$ defined before. Each configuration of the catalog in Fig.~\ref{fig.anal_rules} is obtained with a probability $P[\cdots]$ given by,

           \begin{align}
             P[\circ\circ\circ\circ\circ]&=P_n \nonumber \\
             P[\bullet\circ\circ\circ\circ\circ\circ]&=P[\circ\circ\circ\circ\circ\circ\bullet]=P_{n+1}-P_{n+2}
           \label{eq.anal_mechanism} \\
             P[\circ\circ\circ\circ\circ\circ\circ]&=P_{n+2} \nonumber \\
             P[\bullet\circ\circ\circ\circ\circ\bullet]&=2P_n-2P\mbox[\bullet\circ\circ\circ\circ\circ\circ]-P[\circ\circ\circ\circ\circ\circ\circ] \nonumber \\
             &=P_n-2P_{n+1}+P_{n+2}. \nonumber
           \end{align}
The proper set of rate equations depends on the case. Below, we describe this set for equal rates of dimers and monomers, adsorption of a preferential dimer site, and different rates for dimers and monomers.

\subsection{Equal deposition rate of dimers and monomers}\label{sec:eq_abs}

  For equal deposition rates of monomers and dimers, the rate of both species is considered as $k$. Since $P_n$ refers to the probability that a randomly selected site belongs to any segment of $n$ empty sites, which can be part of a larger one, this probability can never increase with time. The rate of change of $P_n$ due to the adsorption of dimers is given by,
           \begin{align}
             \left(\dot{P}_n\right)_d=&-k(n-1)P_n-kP[\bullet\circ\circ\circ\circ\circ\circ] \nonumber \\
             &-kP[\circ\circ\circ\circ\circ\circ\bullet]-2kP[\circ\circ\circ\circ\circ\circ\circ] \label{eq.Pn_dimers_eqrates}  \\
             =&-(n-1)kP_n-2kP_{n+1}, \nonumber
           \end{align}
where $(n-1)$ corresponds to the destruction rate of a segment with, at least, $n$ empty sites, which is zero for $n=1$, see Fig.~\ref{fig.anal_rules}(a). The rate of change due to monomers adsorption is,
           \begin{align}
             \left(\dot{P}_n\right)_m=&-2kP[\bullet\circ\circ\circ\circ\circ\bullet]-kP[\bullet\circ\circ\circ\circ\circ\circ] \nonumber \\
             &-kP[\circ\circ\circ\circ\circ\circ\bullet]\label{eq.Pn_monomers_eqrates} \\
             =&-2k(P_n-P_{n+1}). \nonumber
           \end{align}
  From Eqs.~(\ref{eq.Pn_dimers_eqrates})~and~(\ref{eq.Pn_monomers_eqrates}), the total rate of change is given by,
           \begin{align}
             \left(\dot{P}_n\right)_T=\left(\dot{P}_n\right)_d+\left(\dot{P}_n\right)_m=-(n+1)kP_n.  \label{eq.Pn_total_eqrates}
           \end{align}
  This result is equivalent to consider that, regardless the type of adsorption, a segment of $n$, or more, empty sites can be destroyed by adsorption on $(n+1)$ different places. This equation gives $P_n(t)=\exp{\left[-(n+1)kt\right]}$. The coverage $\theta$ can then be obtained from the probability that a certain site is part of a segment of size $n\geq1$, i.e.,
	   \begin{align}
             \theta(t)=1-P_1(t)=1-\exp{(-2kt)}. \label{eq.coverage_eqrates}
	   \end{align}
  Defining $s_m(n)$ as the rate of monomers adsorption on a $n$-segment, i.e., $s_m(n)=-\left(\dot{P}_n\right)_m$, we obtain,
           \begin{align}
             s_m(n)&=2k(P_n-P_{n+1}) \label{eq.sm_eq1} \\
             &=2k\left[\exp{(-\left[n+1\right]kt)}-\exp{(-\left[n+2\right]kt)}\right].\nonumber
           \end{align}
  The adsorption of monomers on a segment of size $n\geq1$, is given by,
	   \begin{equation}
             s_m(n=1)=2k\left[\exp{(-2kt)}-\exp{(-3kt)}\right].
             \label{eq.sm_n1}
	   \end{equation}
  From the rate of adsorption, the coverage of monomers can be obtained from $\dot{\theta}_m=s_m$ for $n=1$ giving,
           \begin{equation}
             \theta_m(t)=\left[1-\exp{(-2kt)}\right]+\frac{2}{3}\left[\exp{(-3kt)}-1\right].
           \label{eq.thetam_eqrates2}
           \end{equation}
  Defining now $s_d(n)$ as the rate of dimers adsorption on a $n$-segment, i.e., $s_d(n)=-\left(\dot{P}_n\right)_d$, we obtain,
           \begin{align}
             s_d(n)=&k[(n-1)\exp{(-\left[n+1\right]kt)} \label{eq.sd_eq1} \\
                    &+2\exp{(-\left[n+2\right]kt)}]. \nonumber
           \end{align}
  The adsorption of dimers on a segment of size $n\geq1$, is then given by,
	   \begin{equation}
             s_d(n=1)=2k\exp{(-3kt)}.
             \label{eq.sd_n1}
	   \end{equation}
  Knowing the rate of adsorption, the coverage of dimers can be obtained from the rate equation, $\dot{\theta}_d=s_d(n=1)$,
           \begin{equation}
             \theta_d(t)=\frac{2}{3}\left[1-\exp{(-3kt)}\right].
           \label{eq.thetad_eqrates2}
           \end{equation}

\subsection{Adsorption of a preferential dimer site}\label{sec:pref_carb}

  Considering a preferential dimer site means that the symmetry is broken and the first adsorption on the substrate occurs through a specific compound of the dimer. The cleavage only takes place if there is a neighboring empty site, in any direction, to adsorb the other compound. In 1D, the left site of the dimer is considered as the preferred one. By symmetry, results are independent on the considered one. For this special case, the change on the $P_n$ by dimers is,
           \begin{align}
             \left(\dot{P}_n\right)_d=&-k(n-1)P_n-kP[\bullet\circ\circ\circ\circ\circ\circ] \nonumber\\
            &-kP[\circ\circ\circ\circ\circ\circ\bullet]-2kP[\circ\circ\circ\circ\circ\circ\circ]
           \label{eq.Pn_dimers_leftcarb} \\
             =&-(n-1)kP_n-2kP_{n+1}, \nonumber
           \end{align}
while by monomers is,
           \begin{align}
             \left(\dot{P}_n\right)_m&=-kP[\bullet\circ\circ\circ\circ\circ\bullet]-kP[\circ\circ\circ\circ\circ\circ\bullet]\nonumber\\
             &=-k(P_n-P_{n+1}). \label{eq.Pn_monomers_leftcarb}
           \end{align}
  For the total change on $P_n$ we obtain,
           \begin{align}
             \left(\dot{P}_n\right)_T=\left(\dot{P}_n\right)_d+\left(\dot{P}_n\right)_m=-nkP_n-kP_{n+1}.  \label{eq.Pn_total_leftcarb}
           \end{align}
  Considering the relation between $P_n$ and $P_{n+1}$ as $P_{n+1}=Q_nP_n$ \cite{Evans1997}, then
           \begin{align}
             \dot{P}_{n+1}=-(n+1)kQ_nP_n-kQ_{n+1}Q_nP_n, \label{eq.Pn_total_leftcarb2}
           \end{align}
and plugging it back into Eq.~(\ref{eq.Pn_total_leftcarb}) we obtain,
           \begin{align}
             \frac{dQ_n}{dt}\frac{1}{Q_n}=-(n+1)k+nk-k(Q_{n+1}-Q_n). \label{eq.Pn_total_leftcarb3}
           \end{align}
If we assume $Q_{n+1}=Q_n$, then $\frac{dQ_n}{Q_n}=-kdt$. Therefore, $Q_n(t)=\exp{(-kt)}$, which when replaced in Eq.~(\ref{eq.Pn_total_leftcarb3}) gives $\dot{P}_n=-\left[nk+k\exp{(-kt)}\right]P_n$ and so,
           \begin{equation}
             P_n(t)=\exp{\left[-nkt+\left(\exp{\left[-kt\right]}-1\right)\right]}.
             \label{eq.Pn_total_leftcarb7}
           \end{equation}
  Since the coverage $\theta$ is dependent on the evolution of the probability of finding a segment of size $n\geq1$,
	   \begin{align}
             \theta(t)&=1-P_1(t) \label{eq.coverage_leftcarb} \\
             &=1-\exp{\left[-kt+\left(\exp{\left[-kt\right]}-1\right)\right]}. \nonumber
	   \end{align}
The independent rates of adsorption for dimers and monomers and the subsequent calculation of the coverage for each species are obtained as before.

\subsection{Different deposition rates for dimers and monomers}\label{sec:diff_rates}

  To attempt a generic solution for the rules given by Eq.~(\ref{eq.adsorption_mechanism}), it is necessary to consider different deposition rates for monomers ($k_m$) and dimers ($k_d$). Accounting for the rate of change of $P_n$ by dimers given by Eq.~(\ref{eq.Pn_dimers_eqrates}),
  \begin{equation}
     \left(\dot{P}_n\right)_d=-(n-1)k_dP_n-2k_dP_{n+1},
     \label{eq.Pn_dimers_difrates}
  \end{equation}
while by monomers,
  \begin{equation}
     \left(\dot{P}_n\right)_m=-2k_m(P_n-P_{n+1}).
     \label{eq.Pn_monomers_difrates}
  \end{equation}
  The change on the the total $P_n$ over time is then given by,
           \begin{align}
             \left(\dot{P}_n\right)_T&=\left(\dot{P}_n\right)_d+\left(\dot{P}_n\right)_m
           \label{eq.Pn_total_difrates} \\
             &=-\left[k_d(n-1)+2k_m\right]P_n-2(k_d-k_m)P_{n+1}. \nonumber
           \end{align}
  For the sake of simplicity, we define $\alpha_n=(n-1)k_d+2k_m$ and $\beta=k_d-k_m$. In the same way as before, applying the relation $P_{n+1}=Q_nP_n$, the rate equation for $P_{n+1}$ is,
           \begin{align}
             \dot{P}_{n+1}&=\dot{Q}_nP_n+Q_n\dot{P}_n \label{eq.Pn_total_difrates2} \\
             &=-\alpha_{n+1}Q_nP_n-2\beta Q_{n+1}Q_nP_n, \nonumber
           \end{align}
and replacing $\dot{P_n}$ by Eq.~(\ref{eq.Pn_monomers_difrates}),
           \begin{align}
             &\dot{Q}_nP_n+Q_n\left[-(\alpha_n+2\beta Q_n)P_n\right]= \nonumber \\ 
             &-\left(\alpha_{n+1}Q_n+2\beta  Q_{n+1}Q_n\right)P_n
             \label{eq.Pn_total_difrates3} \\
             &\dot{Q}_n=-k_dQ_n-2\beta(Q_{n+1}-Q_n)Q_n. \nonumber
           \end{align}
Considering $Q_{n+1}=Q_n$ then $Q_n(t)=\exp{(-k_dt)}$. From the above result, Eq.~(\ref{eq.Pn_total_difrates}) simplifies as $\dot{P}_n=-\left(\alpha_n-2\beta Q_n\right)P_n$, which gives,
           \begin{equation}
\label{eq.Pn_total_difrates6}
           \begin{split}
             P_n(t)=&\exp\biggl[-(\left[n-1\right]k_d+2k_m)t\biggr. \\
             &\left.-\frac{2(k_d-k_m)}{k_d}\left(1-\exp{\left[-k_dt\right]}\right)\right].
           \end{split}
           \end{equation}
From Eq.~(\ref{eq.coverage_eqrates}),
	   \begin{align}
           \label{eq.coverage_difrates}
           \begin{split}
             \theta(t)=&1-P_1(t) \\
	     =&1-\exp\biggl[-2k_mt\biggr.\\
             &\biggl.-\frac{2(k_d-k_m)}{k_d}\left(1-\exp{\left[-k_dt\right]}\right)\biggr],
           \end{split}
	   \end{align}
which for $k_m=k_d=k$ boils down to Eq.~(\ref{eq.coverage_eqrates}). If $s_m(n)$ is defined as the rate of monomers adsorption on a $n$-segment $s_m(n)=-\left(\dot{P}_n\right)_m$ and so
           \begin{align}
             s_m(n)=2k_m(1-Q_n)P_n. \label{eq.sm_eq1_difrates}
           \end{align}
  The relations $y=\exp{(-k_dt)}$ and $\gamma=\frac{k_m}{k_d}$ are considered and the adsorption of monomers on a segment of size $n\geq1$ is given by,
	   \begin{equation}
             s_m(1)=2k_m\left(1-y\right)y^{2\gamma}\exp{\left[-2(1-\gamma)(1-y)\right]}.
             \label{eq.sm_n1_difrates}
	   \end{equation}
  From the rate of adsorption, the coverage of monomers is given by $\dot{\theta}_m=s_m$ for $n=1$, and so,
           \begin{equation}
             \theta_m=-\int^{\exp{(-k_dt)}}_1s_m(1)\left(yk_d\right)^{-1}dy.
             \label{eq.thetam_difrates3}
           \end{equation}
  The dimers rate of adsorption is then,
           \begin{align}
             s_d(n)=k_d(n-1)P_n+2k_dQ_nP_n. \label{eq.sd_eq1_difrates}
           \end{align}
  For the sake of simplicity, the relation $y=\exp{(-t)}$ is used, and the adsorption of dimers on a segment of size $n\geq1$ is given by,
	   \begin{equation}
             s_d(1)=2k_dy^{k_d+2k_m}\exp{\left[-\frac{2\left(k_d-k_m\right)}{k_d}\left(1-y^{k_d}\right)\right]}.
             \label{eq.sd_n1_difrates}
	   \end{equation}
  For the rate of adsorption, the coverage of monomers can be given by, $\dot{\theta}_d=s_d$ for $n=1$, which by integrating over $y$ gives,
          \begin{equation}
             \theta_d=-\int^{\exp{(-t)}}_1s_d(1)y^{-1}dy. \label{eq.thetad_difrates3}
           \end{equation}

\section{1D Monte Carlo Simulations}\label{sec:MC_1d}
  We numerically studied the proposed model through Monte Carlo simulations, performed on a lattice with $10^4$ sites, where periodic boundary conditions have been applied  and results have been averaged over $10^2$ samples.

   \begin{figure}[t]
   \begin{center}
   \begin{tabular}{cc}
    \includegraphics[width=4cm]{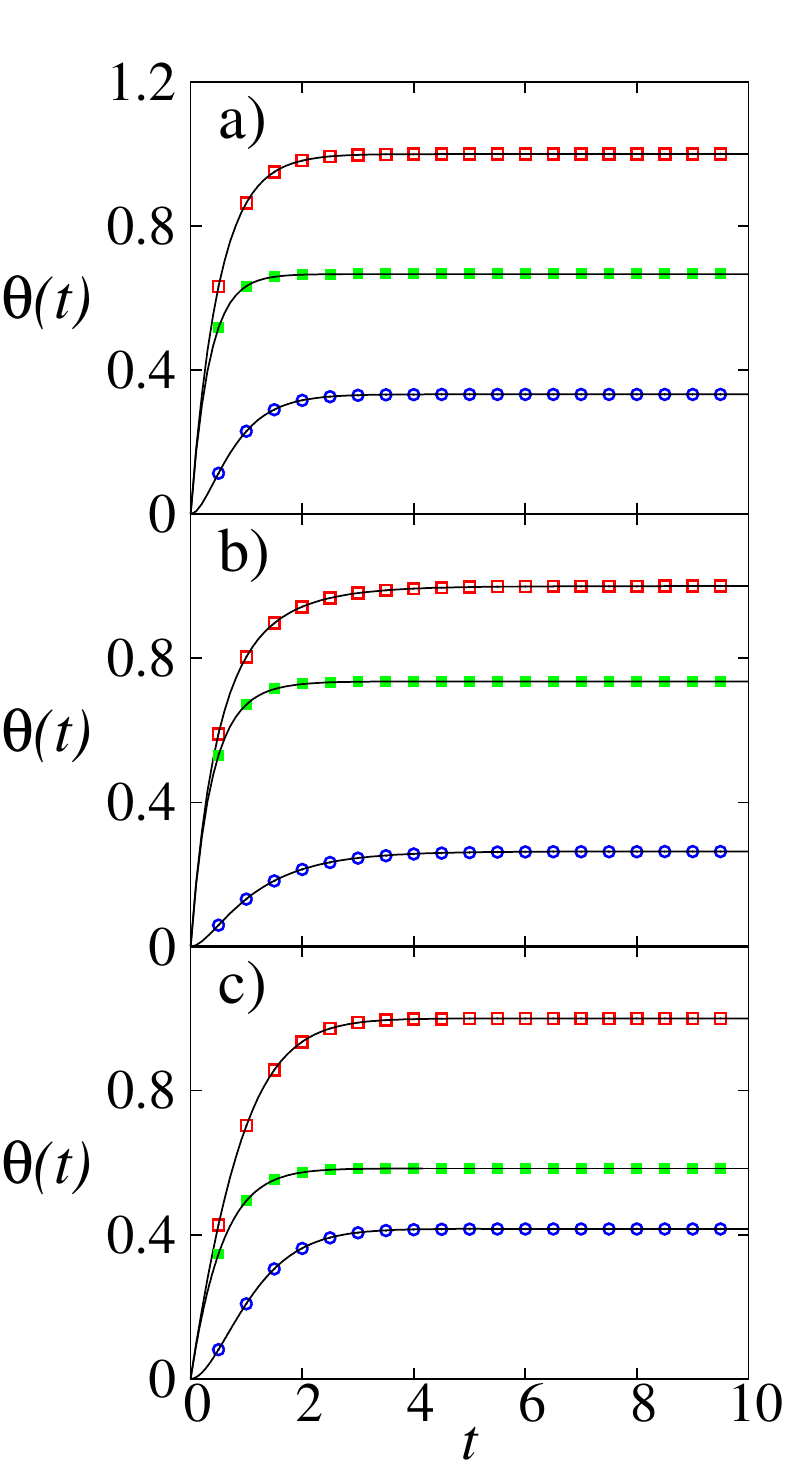} & \includegraphics[width=4cm]{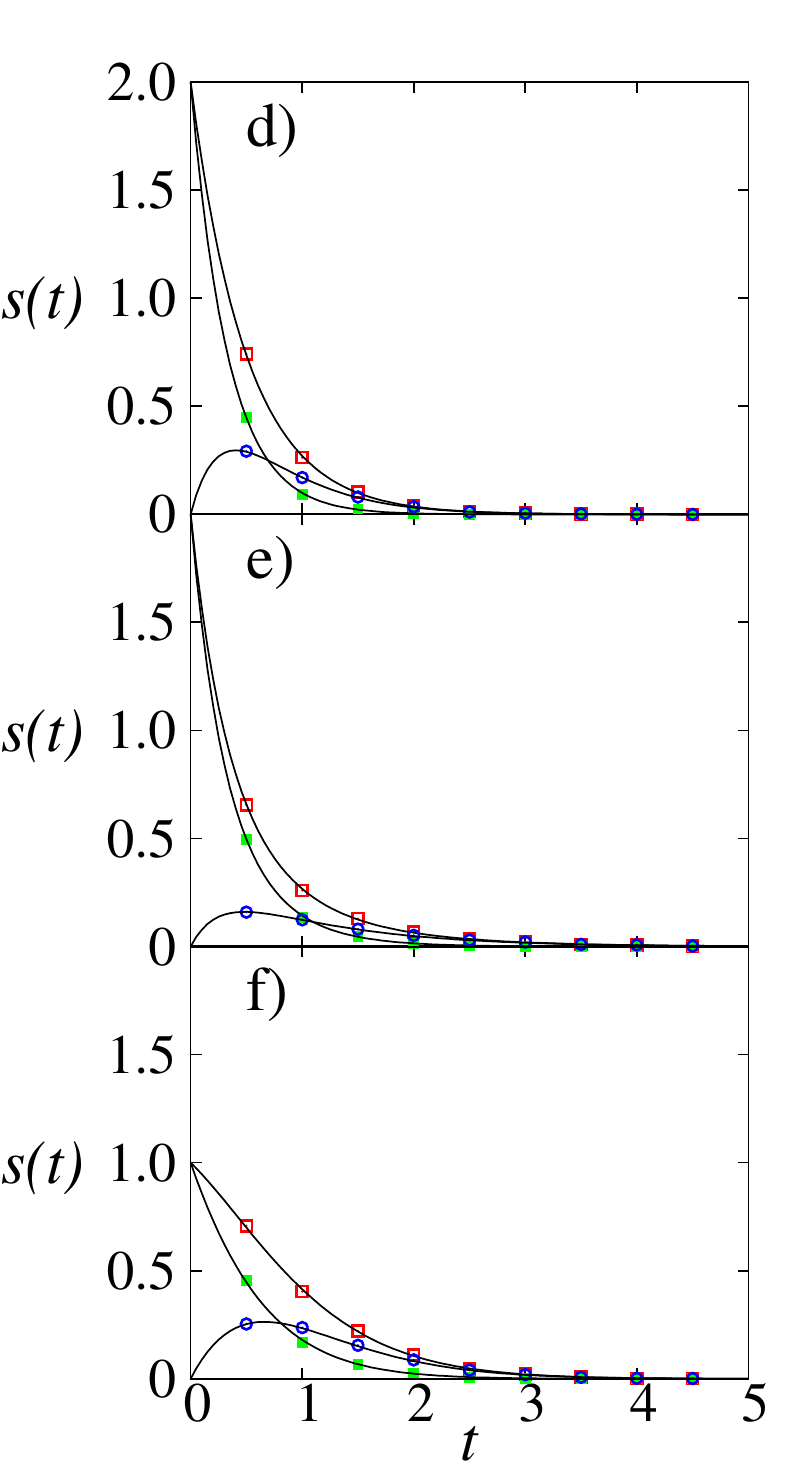}\\
   \end{tabular}
    \caption{Coverage $\theta(t)$ and rates of adsorption $s(t)$ as a function of time obtained through Monte Carlo simulations (symbols) and analytically (solid line) for equal rates of adsorption (a) and (d),  preferential carbon adsorption (b) and (e), and different rates of adsorption ($k_d=0.5$ and $k_m=1$) (c) and (f). Total coverage and rates of adsorption (open squares), dimers (full squares), and monomers (open circles).}
    \label{fig.coverage_eqrates}
   \end{center}
   \end{figure}

  The coverage as a function of time is plotted on Figs.~\ref{fig.coverage_eqrates}(a),~(b),~and~(c) for the three previously described cases. The total coverage $\theta_T$, the coverage of dimers $\theta_d$, and the coverage of monomers $\theta_m$ are computed over 10 Monte Carlo (MC) sweeps, where one MC sweep corresponds to one adsorption attempt per lattice site. Figure~\ref{fig.coverage_eqrates}(a) shows the case of equal deposition rates for dimers and monomers, where $k_d=k_m=1$. The solid lines in the plot represent the analytical solution given by Eqs.~(\ref{eq.coverage_eqrates}),~(\ref{eq.thetam_eqrates2}),~and~(\ref{eq.thetad_eqrates2}). The preferential dimer site rule of adsorption is in Fig.~\ref{fig.coverage_eqrates}(b), where despite the values of the deposition rates being given by $k_d=k_m=1$, the results are equivalent to the ones for $k_d=1$ and $k_m=0.5$, since the deposition of a monomer by the non-favored dimer site is not allowed. For these rules of deposition, the exact results are given by Eqs.~(\ref{eq.coverage_leftcarb}),~(\ref{eq.Pn_dimers_leftcarb}),~and~(\ref{eq.Pn_monomers_leftcarb}). The final case, where a different deposition rate for dimers and monomers is considered, is plotted in Fig.~\ref{fig.coverage_eqrates}(c), with deposition rates of $k_d=0.5$ and $k_m=1$. The exact solution is given by Eqs.~(\ref{eq.coverage_difrates}),~(\ref{eq.thetam_difrates3}),~and~(\ref{eq.thetad_difrates3}). For all cases, data points from Monte Carlo lay on the line given by the exact solution.

  Figures~\ref{fig.coverage_eqrates}~(d),~(e),~and~(f) show the rates of adsorption as a function of time (only five MC sweeps are shown). The plots (d), (e), and (f) correspond, respectively, to equal deposition rates, preferential dimer site rule, and different deposition rates. Under the same conditions as for the coverage study, some particular aspects can be observed. The dimers rate of adsorption, for instance, starts at a value of two for $k_d=1$ since dimers occupy two sites at each deposition. The monomers rate of adsorption starts at zero, and increases as it requires previously adsorption of, at least, one particle. The rate of monomers adsorption increases due to the large influence of the substrate coverage and reaches a maximum when the number of isolated empty sites start to decrease. Exact results are shown for each plot, consistent with the ones obtained with Monte Carlo simulations.

\begin{figure}[t]
   \begin{center}
    \includegraphics[width=8cm]{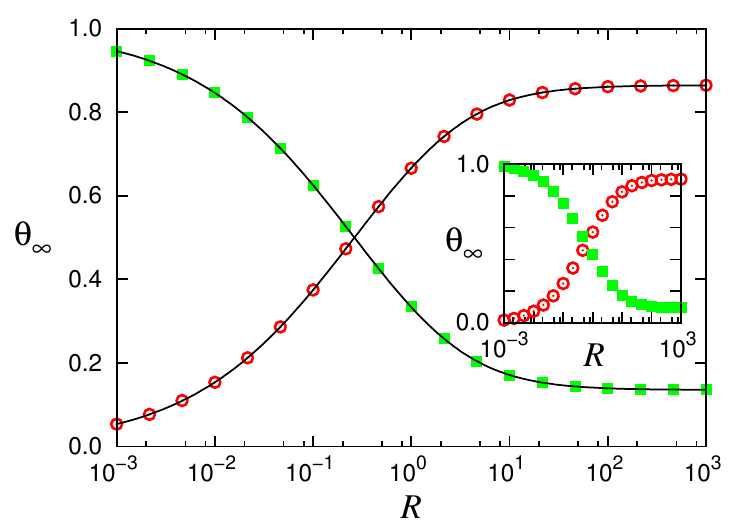}\\
    \caption{Coverage of dimers (open circles) and monomers (full squares) for Monte Carlo simulations and obtained analytically (solid line), with the analytical solution (solid line), at the jamming limit as a function of the ratio between the rate of deposition of dimers and monomers $R$ for the one-dimensional case (two dimensions Monte Carlo results on the inset).}
    \label{fig.coverage_varrates}
   \end{center}
   \end{figure}

  Since desorption is neglected, a jamming limit is obtained where no further particles can be adsorbed. In Fig.~\ref{fig.coverage_varrates}, we see the coverage in the jamming limit $\theta_{\infty}$ as a function of the ratio between dimers and monomers deposition rates, $R=k_d/k_m$, where the solid line is the analytical solution, open circles are dimers, and full squares monomers. It can be observed that in the limit of $R\ll1$, a complete coverage of monomers is found, except for one dimer that always need to be adsorbed to start the monomers deposition. With the decrease in the coverage of monomers an increase in the coverage of dimers is observed, where equal coverage is reached for a ratio $R=0.207\pm0.006$. In the limit of $R\gg1$, a maximum coverage of dimers is obtained in agreement with the classical adsorption of dimers in a one-dimensional lattice \cite{Evans1993a}. Monte Carlo results are also in agreement with the analytical solution.

\section{2D Monte Carlo Simulations}\label{sec:2d_results}

   \begin{figure*}[t]
   \begin{center}
   \begin{tabular}{cc}
    \includegraphics[width=8cm]{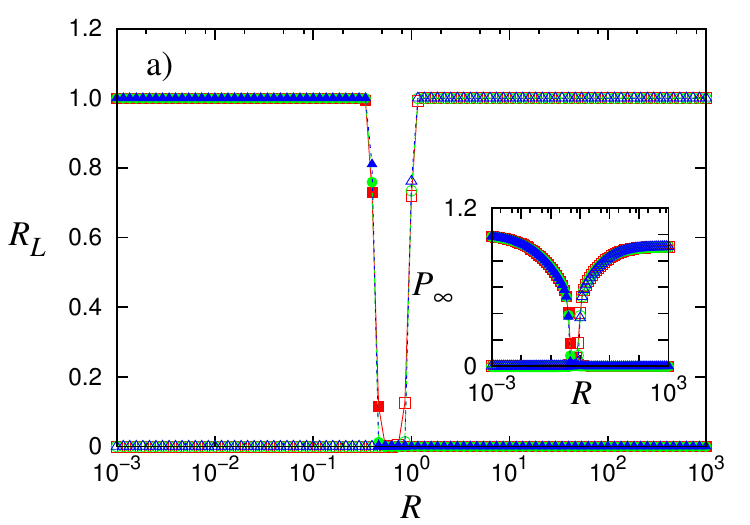} &  \includegraphics[width=8cm]{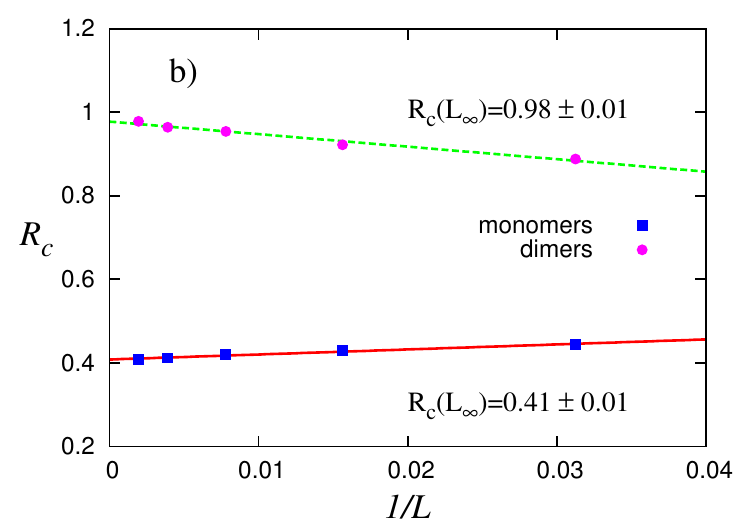} \\
   \end{tabular}
    \caption{a) Dependency on the ratio $R$ of the spanning probability $R_L$ for  dimers (open) and monomers (full). Square lattices have been considered with $128^2$ (squares), $256^2$ (circles), and $512^2$ (triangles) lattice sites, for the spanning probability $R_L$ and fraction of sites belonging to the largest cluster $P_\infty$ (inset). b) Percolation threshold ($R_c$) as a function of the system sizes, for linear sizes of $L=\{32,64,128,256,512\}$, for dimers (full circles) and monomers (full squares).}
    \label{fig.percolation_clean}
   \end{center}
   \end{figure*}

  \begin{figure*}[t]
   \begin{center}
   \begin{tabular}{cc}
    \includegraphics[width=8cm]{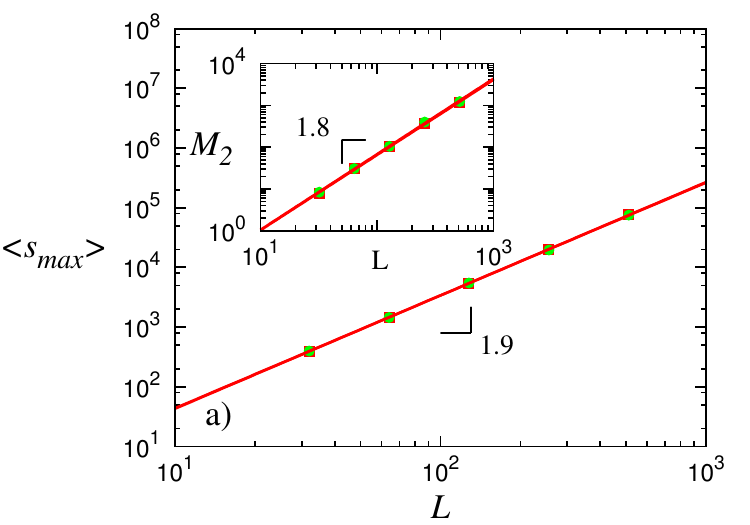} & \includegraphics[width=8cm]{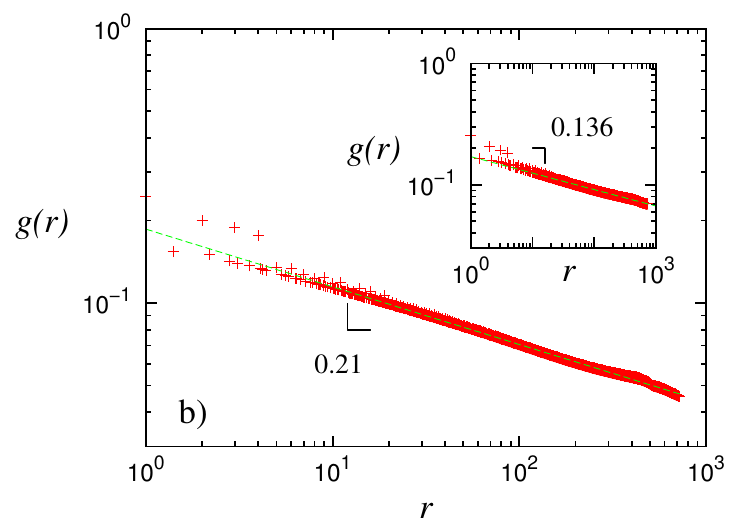}\\
   \end{tabular}
    \caption{a) Largest cluster $s_{max}$ and (inset) second moment of the cluster size distribution function $M_2$ at $R_c$ as a function of the system sizes, for linear sizes of $L=\{32, 64, 128, 256, 512\}$, for dimers (full circles), and monomers (full squares). Fractal dimension for monomers and dimers of $D_m=1.898\pm 0.008$ and $D_m=1.890\pm 0.009$ respectively. b) correlation function $g(r)$ for dimers (monomers on the inset) for a system of linear size $L=1024$ and averaged over 100 samples with power-law exponent of $\eta=0.2101\pm 0.0002$ ($\eta=0.1361\pm 0.0002$).}
    \label{fig.percolation_pinf}
   \end{center}
   \end{figure*}

 In two dimensions, even for the simplest case of dimer adsorption, no analytical solution have been found. However, it is a case of interest, specially the regular square lattice which reproduces, for example, the topology of the Pt(100) surface. In this section, we study the proposed model on a square lattice through Monte Carlo simulations. We devote special attention to the percolation properties of aggregates of monomers and dimers \cite{Rampf2002}.

 Monte Carlo simulations have been performed on square lattices of linear sizes $L=\{128,256,512\}$ in units of lattice sites, with periodic boundary conditions in both directions. Results have been averaged over $10^4$ samples. To decrease the computational effort, a rejection free algorithm was implemented, where the next adsorption trial takes place on an empty site randomly selected from a list of available sites, where the weight of each configuration is properly taken into account. To accurately follow the time evolution, the entire population of events is considered as well as the rate of monomers and dimers adsorption.

 Simulations have been performed on both clean and impurities-covered substrates. Impurities are considered quenched and to solely influence the adsorption process by purely geometrical restrictions.

 \subsection{Clean Substrate}\label{sec:clean}
  On a clean substrate, the coverage of dimers is larger than the coverage of monomers and the rate of adsorption of monomers as a function of time has a maximum, as also observed in the one-dimensional case. However, the coverage of monomers is favored in two dimensions since each deposited dimer has a greater influence on monomer deposition than in one dimension, mainly due to a larger fraction of configurations with occupied and empty neighboring sites. This can be observed on the inset of Fig.~\ref{fig.coverage_varrates}, where we plot the jamming limit $\theta_{\infty}$ as a function of the ratio $R$. In this case, the point of equal coverage for dimers and monomers occurs for a ratio $R\approx0.6$, which is larger than in one dimension, corroborating that monomers are favored.

 \begin{figure}[t]
   \begin{center}
    \includegraphics[width=8cm]{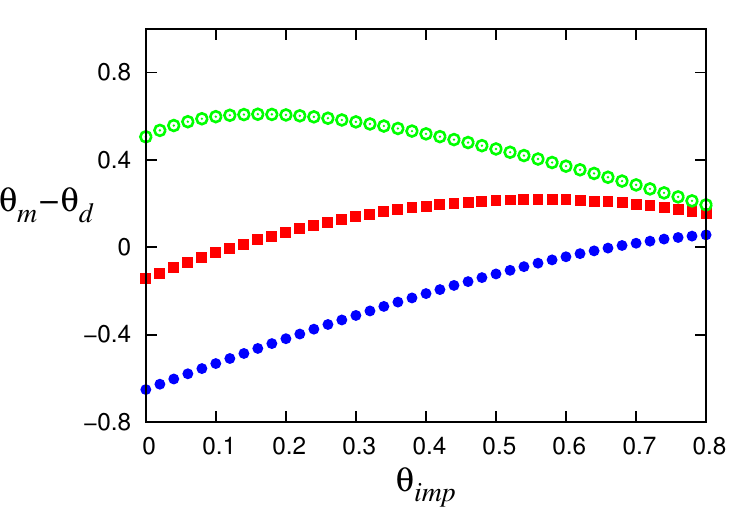}\\
    \caption{Difference in the jamming coverage of monomers and dimers as a function of the impurities coverage, in 2D, for $R=0.1$ (open circles), $R=1$ (full squares), and $R=10$ (full circles).}
    \label{fig.coverage_diff_varrates_2d}
   \end{center}
   \end{figure}

  The percolation properties are analyzed by identifying clusters with the Hoshen-Kopelman algorithm \cite{Hoshen1976}. While for lower values of R the system is dominated by monomers, as discussed for 1D, for higher values of R dimers dominate. Percolation of monomers or dimers is then observed with R as a control parameter. In Fig.~\ref{fig.percolation_clean}(a) we plot the spanning probability $R_L$ defined as the probability of having a percolation cluster touching opposite borders of the lattice. At the percolation transition of both monomers and dimers, the fraction of sites occupied by the specie under study is compatible with the percolation threshold for site percolation in the considered topology \cite{Stauffer1994}. In the inset of Fig.~\ref{fig.percolation_clean}(a) is the fraction of sites belonging to the largest cluster $P_{\infty}$ (the order parameter of the percolation transition).

  The percolation threshold $R_c$ can be estimated analyzing the maximum value on the standard deviation of the spanning probability. It can be observed in Fig.~\ref{fig.percolation_clean}(b), that the percolation threshold scales linearly with the inverse of the lattice lateral size, L. Obtaining for dimers $R_c (L_{\infty})=0.98\pm0.01$, and for monomers $R_c(L_{\infty})=0.41\pm0.01$.

  The scaling of the average size of the largest cluster $\langle s_{max}\rangle $ at $R_c$ is in Fig.~\ref{fig.percolation_pinf}(a) and scales as $\sim L^{D_f}$, where $D_f$ is the fractal dimension. For both monomers and dimers the obtained scaling for the mass of the largest cluster is consistent with the fractal dimension $D_f=91/48=1.8958$ of the classical percolation universality class \cite{Stauffer1994}. 

  \begin{figure}[t]
   \begin{center}
    \includegraphics[width=8cm]{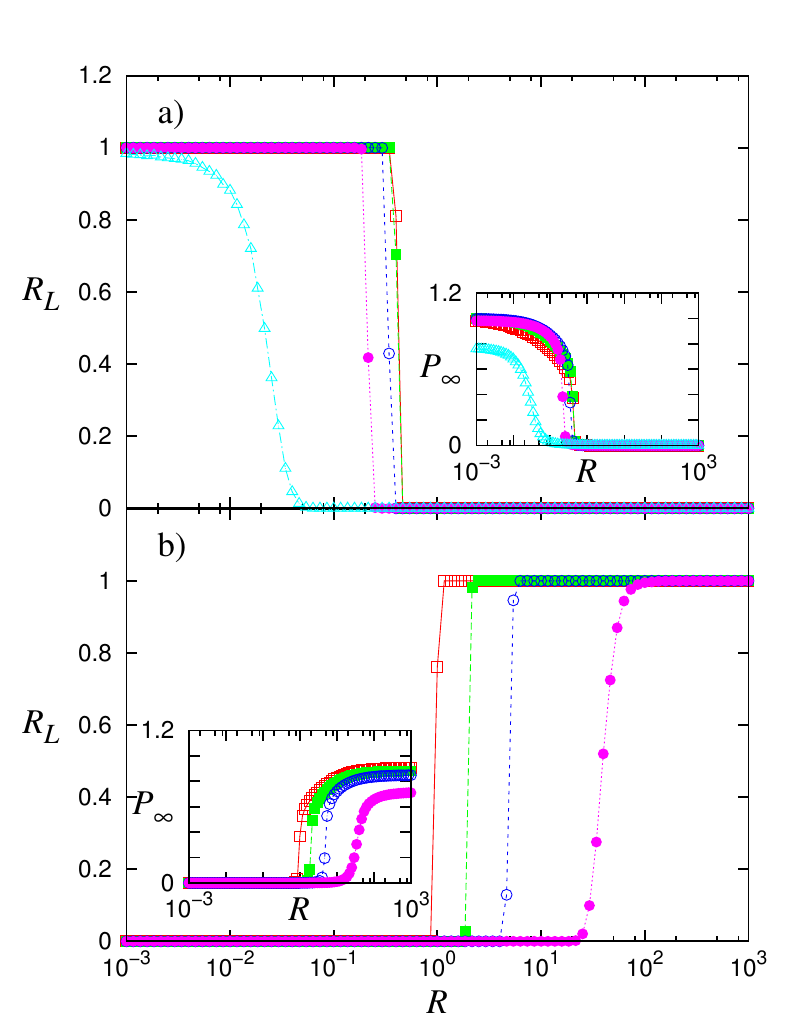} \\
    \caption{Wrapping probability as a function of the ratio $R$, in 2D, for a) monomers with impurities coverage from right to left of 0\%, 10\%, 20\%, 30\% and, 40\%, and b) dimers with impurities coverage from left to right of 0\%, 10\%, 20\%, and 30\%. Simulation of a system size of $512^2$. Fraction of sites belonging to the largest cluster $P_\infty$ on the inset.}
    \label{fig.percolation_imp}
   \end{center}
   \end{figure}

  Another important parameter to be taken into account is the second momentum of the cluster-size distribution given by,
       \begin{equation}
        M_2=\frac{\sum_{i\neq max} s_i^2}{N},
        \label{eq.m2}
       \end{equation}
where $s$ is the cluster size, $N$ is the total number of lattice sites, and the sum runs over all clusters excluding the largest one. The variable $M_2$ at $R_c$ as a function of the system size is in the inset of Fig.~\ref{fig.percolation_pinf}(a), where another scaling behavior is observed consistent with $M_2\sim L^{\frac{\gamma}{\nu}}$, with $\frac{\gamma}{\nu}=1.80\pm 0.02$, in agreement with the scaling relation $\frac{\gamma}{\nu}=2D_f-d$.

  We measured the correlation function, also known as connectivity correlation function, defined as
       \begin{equation}
        g(r)=\langle \delta_{ij}\rangle -\frac{s_{max}^2}{N^2},
        \label{eq.correlation}
       \end{equation}
where $\delta_{ij}$ is 1 if both sites $i$ and $j$ are occupied by the same cluster and zero otherwise and $s_{max}$ is the size of the largest cluster. Figure~\ref{fig.percolation_pinf}(b) shows that at $R_c$ both correlation functions are power laws with an exponent consistent with the one for random percolation \cite{Stauffer1994}.

 \subsection{Substrate with impurities}\label{sec:impurity}

 To account for the presence of impurities (e.g., Pb atoms \cite{Spencer1983}), quenched impurities are randomly distributed on the substrate. These impurities do not react and remain immobile, influencing only the adsorption, as an occupied site, which promotes the adsorption of monomers.
 
 Since impurities geometrically favor the coverage of monomers, the value of $\theta_m-\theta_d$ as a function of impurities coverage $\theta_{imp}$ is plotted in Fig.~\ref{fig.coverage_diff_varrates_2d}. A maximum is observed at a specific value of the coverage by impurities. The position at which the maximum occurs depends on the ratio $R$; low values of $R$ favor monomers leading to an earlier maximum while a high ratio disfavors monomers, thus, shifting the position of the maximum to larger values. These results open up the possibility of tuning the production of monomers by controlling the fraction of impurities.

 Additionally, impurities also shift the percolation threshold. Figures~\ref{fig.percolation_imp}(a)~and~(b), show the spanning probability $R_L$ as a function of the ratio R for different values of impurities coverage. The monomers percolation transition is mainly affected at larger impurities coverage and is shifted to lower values of $R$. The dimer percolation transition, on the other hand, is shifted to higher values of $R$. At higher values of impurities coverage, the clusters of impurities predominate on the surface and neither monomers nor dimers percolate. The insets of Figs.~\ref{fig.percolation_imp}(a)~and~(b) show the fraction of sites belonging to the largest cluster $P_{\infty}$ as a function of $R$, with $P_{\infty}=\frac{\langle s_{max}\rangle }{N\left(1-\theta_{imp}\right)}$, where $N$ is the total number of lattice sites, and $\theta_{imp}$ is the coverage of impurities. In the case of monomers, as disclosed by the behavior of the spanning probability, the size of the largest cluster is only significantly affected by impurities for values of impurities coverage above 30\%. In the case of dimers, impurities have a larger effect on $R_L$ and $P_{\infty}$.

\section{Final remarks}\label{sec:conclusion}

  We introduced a model based on random sequential adsorption of monomers and dimers, representing, acetyl and cleaved products, respectively, in the low desorption limit.
  The kinetic rules based on recent results by Wang and Liu are dependent on the local configuration provided by the neighboring adsorbed sites instead of configuration-independent rates.

  The properties of the model were studied in the 1D lattice and also extended to a 2D square lattice. In the latter case, the model describes the mechanisms of ethanol electro-oxidation on a Pt(100) surface, suggested by Wang and Liu \cite{Wang2008a}. 
  In 1D, we have analytically solved the model in three different cases: same deposition rate for dimer and monomer adsorption, preferential dimer site adsorption, and different deposition rates. Monte Carlo simulations are in agreement with the analytical solution. 
  In 2D we have studied the jammed state and percolation transition through Monte Carlo simulations. The percolation properties of the adsorbed species reveal that, while monomers percolate at low ratios of dimers/monomers deposition rate, dimers percolate at high ratios. The influence of impurities has also been monitored, disclosing that the coverage of monomers is significantly improved by their presence.

  In the present work, we restricted ourselves to study a system as simplified as proposed in the Introduction. One can clearly devise extensions to the basic model to include desorption and reaction pathways not included in the present paper.
  Molecular dynamics studies could, in principle, provide a more complete picture of the particles arrangements on the surface, but at a shorter timescale.
  It would certainly be interesting to study the atomistic mechanisms based on cooperative thermal effects that could affect some of the reaction pathways.
  
  \acknowledgments{
  The authors are thankful to Fundac\~ao para a Ci\^encia e a Tecnologia fellowships (CD SFRH/BD/31833/2006, AC SFRH/BPD/3475/2007) and the warm hospitality of the T-1 group at the Los Alamos National Laboratory during the tenure of the fellowships. Work at Los Alamos National Laboratory (LANL) was supported by United States Department of Energy (U.S. DOE). LANL is operated by Los Alamos National Security, LLC for National Nuclear Security Administration of U.S. DOE under Contract No. DE-AC52-06NA25396.
  AC acknowledges useful comments by Neil Henson on the early stages of this work.}

  \bibliographystyle{apsrev}
  \bibliography{clean_Ethanol}

\end{document}